# Comprehensive understanding of water-driven graphene wrinkle life-cycle towards applications in flexible electronics: A computational study

Jatin Kashyap[1], Eui-Hyeok Yang[2], Dibakar Datta[*,1]

[1] Department of Mechanical and Industrial Engineering, New Jersey Institute of Technology, Newark, NJ 07103, USA
[2] Department of Mechanical Engineering, Stevens Institute of Technology, Hoboken, NJ 07030, USA

*Corresponding Author (Email: dibakar.datta@njit.edu)

**Abstract**

Thin materials are known to wrinkle under thermodynamic perturbations, which results in non-linearity in associated physics. The presence of wrinkles in Graphene Nanoribbons (GNR) and other two-dimensional (2D) materials significantly alter their mechanical, electronic, optical properties, which can be either beneficial or detrimental. Experimentally, it has been observed that during the commonly used growth process of GNR, water molecules, sourced from ambient humidity, can be diffused in between GNR and the substrate. The water diffusion causes wrinkle formation in GNR, which influences its properties. Furthermore, the diffused water eventually dries, creating the alteration not only in the geometry of Wrinkled Graphene Nanoribbons (WGNR) but also its features. Computational analysis of the fundamental mechanisms of formation, evolution, and collapse of wrinkles due to water diffusion and evaporation can provide an atomistic-level understanding of the phenomena. This understanding is essential to give the guidelines to engineer wrinkles in graphene for specific requirements. Therefore, in this work, Molecular Dynamics (MD) simulations are performed to model the water diffusion and evaporation in between GNR and its substrate, and their effect on wrinkle formation and dynamics. Additionally, Density Functional Theory (DFT)-based analysis is used to characterize the difference in the electronic structure of WGNR caused by the change in wrinkle geometry. Our study reveals that the initially distributed wrinkles tend to coalesce to form a localized wrinkle whose configuration depends on the initial wrinkle geometry and the amount of diffused water. The wrinkle configuration changes upon drying, while it remains static until the complete drying. The movement of the localized wrinkle is the combination of three fundamental modes – bending, buckling, and sliding. The stress analysis reveals that the maximum stress is at the base of the wrinkle, and its magnitude is always below the plasticity limit. The DFT results provide insight into the potential of using the wrinkles to control the direction of electron-flow for the applications in flexible electronics. The computational results provide a fundamental understanding of the chemo-electro-nanomechanics of the WGNR.

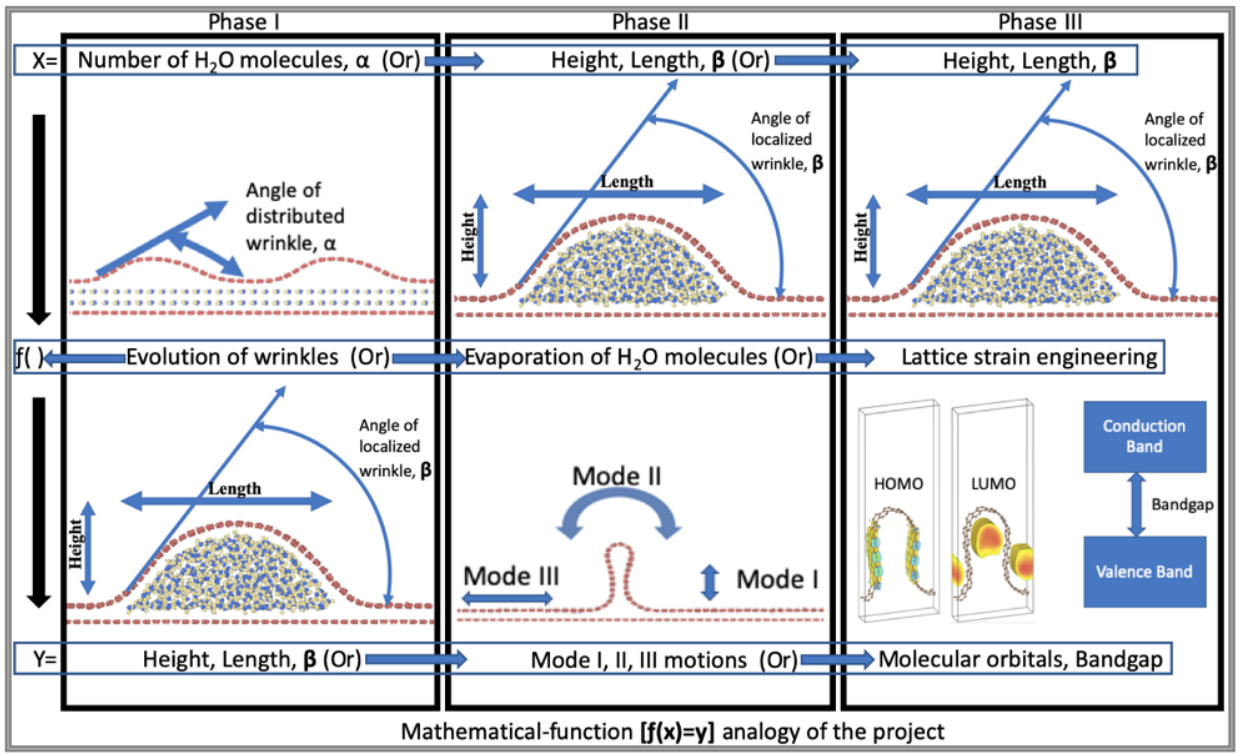

## 1. Introduction

Graphene has been the cynosure in the world of two-dimensional (2D) materials over the last decade [1] owing to its applications in various fields, including electrode materials for energy storage [2], targeted drug delivery [3], environmental protection [4], tissue engineering [5], quantum dots [6], light processing [7], and fuel cells [7]. In contrast to the zero-dimensional (0D) and one-dimensional (1D) materials, 2D materials, including graphene, are less resistant to buckling, and they tend to form surface corrugations such as wrinkles, ripples, crumples [8]. The presence of these surface texture (e.g., wrinkle, graphene-nanobubbles) significantly alters its mechanical, electronic, optical properties for various applications in fields, including electronic [9], sensing and actuation [10], energy storage [11,12], chemical reactions [13-15], and bioelectronics [16]. Depending on applications, wrinkling can be "either boon or a bane". The wrinkling in graphene originates during its different synthesis processes such as mechanical exfoliation [17], layer transfer after chemical vapor deposition [18, 19], reduction of graphite oxides [20], and graphene oxide [21]. During the commonly used synthesis process of graphene in ambient conditions, typically, water molecules in the air are diffused on graphene surface or in between graphene and the underlying substrate. Because of the high surface-to-volume ratio, the diffused water significantly influences the physical properties of graphene [22].

A study by Lee et al. [23] revealed that during the mechanical exfoliation of graphene onto a $SiO_2$ substrate, water molecules diffuse between graphene and substrate, forming ice-like molecular layers when the surface of $SiO_2$ is fully covered (Figure 1). The diffused water molecules and its evaporation cause wrinkles and folding of graphene with the corresponding lattice-strain. The resulting structure is called Wrinkled Graphene Nanoribbons (WGNR). A similar phenomenon was also observed in the case of a mica substrate in their study. Therefore, it is essential to unravel in-depth how exactly water diffusion and evaporation govern the WGNR formation. The atomic-level insight into the mechanisms of wrinkle formation, evolution, and collapse will help us engineer the WGNR for various applications, including flexible electronics.

There are many computational studies on the mechanism of wrinkle formation, evolution, and the collapse of graphene. Molecular Dynamics (MD) simulations were performed to study the wrinkle formation due to tension [24], torsion [25], shear [26], hydrogen functionalization [27], and defects [28] in monolayer graphene. Computational modeling results on the effect of substrate stretching on the wrinkle formation in graphene were reported [29,30]. There are many studies of graphene, such as the vibration characteristics of wrinkles [31], making nanoribbons arrays with wrinkles [32], and utilization of wrinkles for the fabrication of nano-bio devices [33], and the applications of wrinkles on electronics [34]. However, none of these studies considered the effect of water and its evaporation.

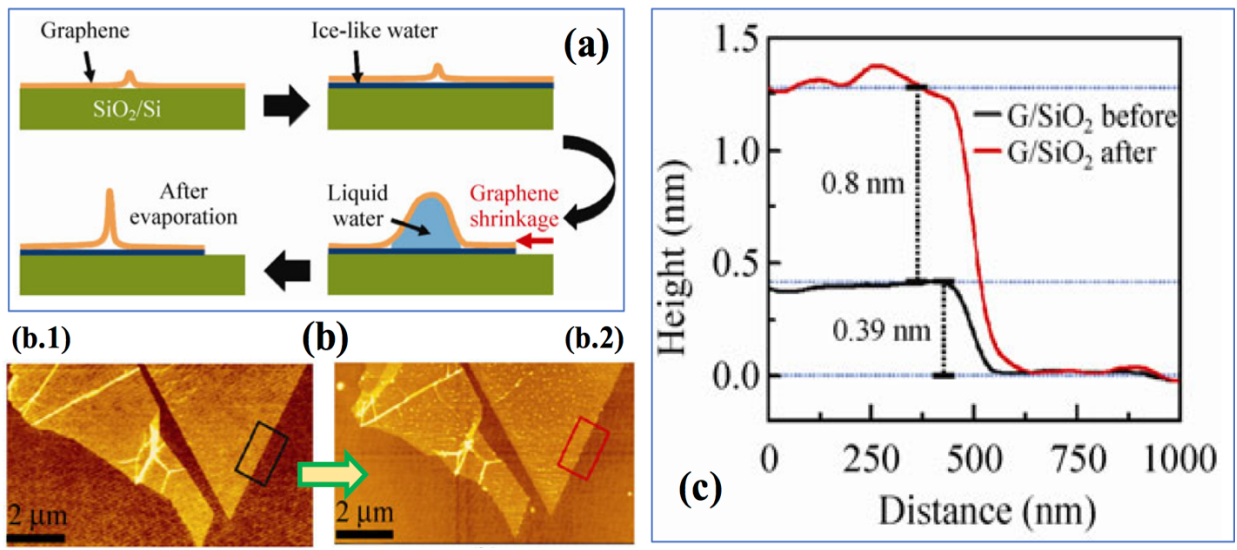

**Figure 1:** (a) Schematic diagram of the wrinkle formation process through the ice-like water formation and liquid water diffusion at high humidity, and liquid water evaporation in a dry environment. (b) Height changes in a graphene/$SiO_2$ structure after exposure to high humidity. AFM topographic images of a monolayer graphene mechanically exfoliated on $SiO_2$ (b.1) before and (b.2) after the high humidity exposure. (c) Averaged height profiles measured for the areas marked with black and red rectangle in (b.1) and (b.2), respectively. G denotes monolayer graphene. **© Copyright : Springer Science + Business Media, Lee et. al. Nano Res. 2012, 5(10):710-717**

The influence of wrinkle on the contact angle (hydrophobicity) of water on the graphene surface was studied by Huang et al. [35]. The self-folding mechanics of surface wettability on patterned graphene nanoribbons by water evaporation was studied by Zhang et al. [36]. We also investigated the interaction of graphene nanoribbons with water droplets [3,37]. The effect of diffused hydrogen molecules in between graphene and copper substrate on the size of wrinkle formation was studied by Wang et al. [38]. The change in the shape of the graphene-nanobubbles under the effect of trapped water was studied by Kalashami et al. [39]. Moreover, there are some studies on the first principle-based analysis of WGNR [40-42]. However, these studies do not address the questions of wrinkle dynamics and its impact on electronic structures due to water diffusion and evaporation. But it is crucial to understand these issues because, in the actual synthesis process, water diffusion and evaporation govern the wrinkle formation and dynamics (Figure 1). To the best of our knowledge, there are no computational studies considering the effect of diffused water and its evaporation on the wrinkle life-cycle and its influence on the electronic properties.

Therefore, several questions are still not addressed: How does the number of diffused water molecules affect the final wrinkle configuration? If graphene already has wrinkles, how does the initial configuration influence the final localized wrinkle configuration? During water evaporation, how does the wrinkle morphology change? Due to generated lattice strains, the stress in planar graphene is substantially different from wrinkled graphene. How much is the maximum stress generated in the graphene with a localized wrinkle? The movement of the wrinkle with diffused water inside is significantly different from the case with no diffused water. How does the diffused water influence the dynamics of wrinkles? How do the electronic properties change with different wrinkle geometry formed due to water diffusion and its evaporation? The atomistic and molecular simulations are essential to address these issues since it is challenging to answer these questions via experimental approaches alone.

In this study, we provide a detailed explanation of water diffusion and its impact on a wrinkled bilayer graphene system characterized by geometrical and electronic parameters. Our results can articulate the reasoning behind various applications based upon the interaction of water with the graphene system [39,40], and change in electronic properties of different 2D materials with wrinkles [41,42]. Our study is composed of three phases. Phase I addresses the wrinkling formation and evolution due to diffused water in between graphene with distributed wrinkles and the underlying supported graphene substrate. We vary the amount of diffused water and the initial angle of the distributed wrinkles in graphene, and study the formation of final localized wrinkle configuration and the stress generated in wrinkled graphene. Phase II deals with the evaporation of diffused water under the localized wrinkle, and the effect of drying on the collapse (structural changes) of the wrinkle, i.e., wrinkle life-cycle. We show that, upon complete drying, the resulting localized wrinkle undergoes motion that is composed of three fundamental modes – bending, buckling, and sliding. Lastly, Phase III considers the free-standing WGNR with four different wrinkle angles and studies its influence on the electronic properties. For Phase I and II, we use the Molecular Dynamics (MD) approach for the bilayer systems, i.e, wrinkled graphene on the graphene substrate. To simplify the model, we consider a fixed graphene substrate instead of a conventional $SiO_2$ or copper substrate. For Phase III, we perform the Density Functional Theory (DFT) calculations to analyze the electronic properties of free-standing WGNR. Overall, our study attempts to provide a comprehensive understanding of the wrinkle life-cycle and its potential applications in flexible electronics.

## 2. Models and Methods

### 2.1 Molecular Dynamics (MD)

This study is divided into *three phases* based on which part of the wrinkle life-cycle is being studied, i.e., formation, evolution, collapse, and the approach used for analysis, i.e., MD or DFT. In the *first phase*, we studied the wrinkle formation and evolution. In this phase, as shown in Figure 2a, using the in-house MATLAB code, we created the initial Graphene with Distributed Wrinkles (GDW) supported by a Flat Graphene (FG) with Periodic Boundary Conditions (PBC) (width = 20 Å). Here, we considered different Initial Angle of Wrinkle (IAW) denoted as $\theta_{IAW}$ ($\theta_{IAW}$ = 6°, 11°, 16°, 21°) to generate GDW with four different configurations. Figure 2a shows a representative case of $\theta_{IAW} = 11°$. While changing $\theta_{IAW}$, we kept the number of carbon (C) atoms in GDW the same, but the size of the FG is reduced. For example, if we take a flat graphene sheet and keep compressing to get wrinkles of different angles, the number of atoms in the sheet remains unchanged, but the overall length of the sheet is reduced. Table S1 of the Supplementary Information (SI) shows the C atoms for different $\theta_{IAW}$. We define a parameter called Carbon Ratio (CR), which is the ratio of the number of C atoms in GDW to that of FG. For $\theta_{IAW}$ of 6°, 11°, 16°, 21°; CR values are 1.017, 1.053, 1.11, 1.20 respectively (Table S1).

The Adaptive Intermolecular Reactive Bond Order (AIREBO) potential was used for the C-C interactions [47]. We performed an energy minimization of this system (Figure 2a) using the conjugate gradient method as implemented in the MD package LAMMPS [48]. Figure 2b shows the minimized/equilibrium configuration of GDW supported by the FG. At this point, we studied wrinkling evolution for four cases – (a) without water (Figure 2b), and (b) two-, (c) four-, and (d) six-layer of water in between the GDW and FG (Figure 2c,d,e). In-house MATLAB code was used to insert the water layer in the wrinkled system shown in Figure 2b. The density of the inserted water molecules, i.e., the number of water molecules per unit volume, is the same for all cases. Table S2 of SI shows the amount of water molecules for different cases.

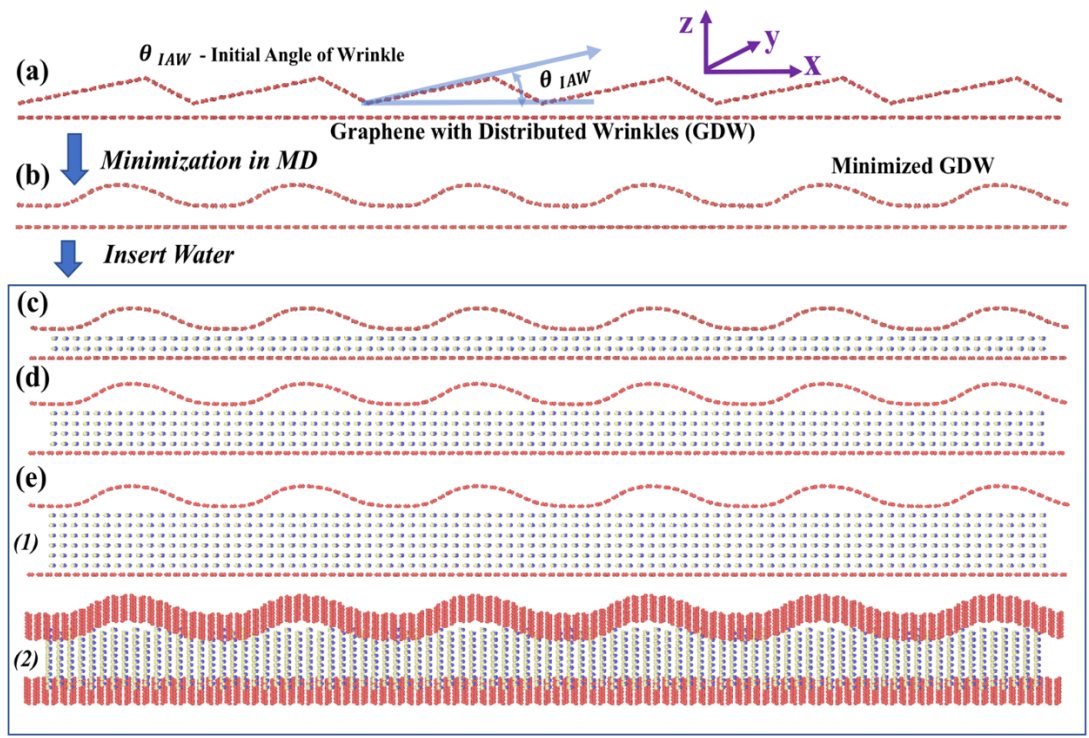

**Figure 2:** (a) V-shaped, hypothetical, wrinkling structure considered for the study (generated using Matlab code). Different Initial Angle of Wrinkling (IAW), $\theta_{IAW}$ = 6°, 11°, 16°, 21°, are considered in this study. (b) Minimization of the system by MD (no water case). (c-e) Insertion of water for further studies. (c) two- (d) four-, and (e) six-layer of water. (e.1) side view and (e.2) inclined view.

The water molecules were added to the system (Figure 2b) in two different ways – (1) While adding the water, the gap between the graphene layers was increased to maintain a fixed distance between the water layers and the adjacent graphene sheet (Figures 2c,d,e). (2) Water layers were added while keeping the distance between GDW and FG to a fixed distance corresponding to the maximum water case considered (i.e., gap shown in Figure 2e is considered for all other cases). However, in both cases, we qualitatively observed the same results. Therefore, we only discuss the results for the cases shown in Figures 2b-e. We used TIP4P potential for water ($H_2O$)[49], and Lennard-Jones type of pair potential for C and $H_2O$ interactions [50]. We used TIP4P potential before for water modeling [3,37], and our computational results were in good agreement with the experimental findings [3].

The methodology of *Phase I* is summarized in the 'Wrinkle Evolution' part of the flow-chart shown in Figure 3. The temperature of the system was controlled at 300 K using the Berendsen thermostat [51]. We then performed MD simulation using NVE ensemble with the timestep of 1 femtosecond. Sufficient MD steps (at least one million steps) were performed to make sure the system reached its lowest energy configurations (discussed later in detail). After the end of the first phase simulations, we stored the system configuration and reused it as the initial structure for the next phase of the project, i.e., water drying case. We performed the stress analysis of the final stabilized graphene sheet with a localized wrinkle using the in-built LAMMPS stress computation algorithm [52]. The detail of the stress calculation methods is provided in the supplementary information (section D).

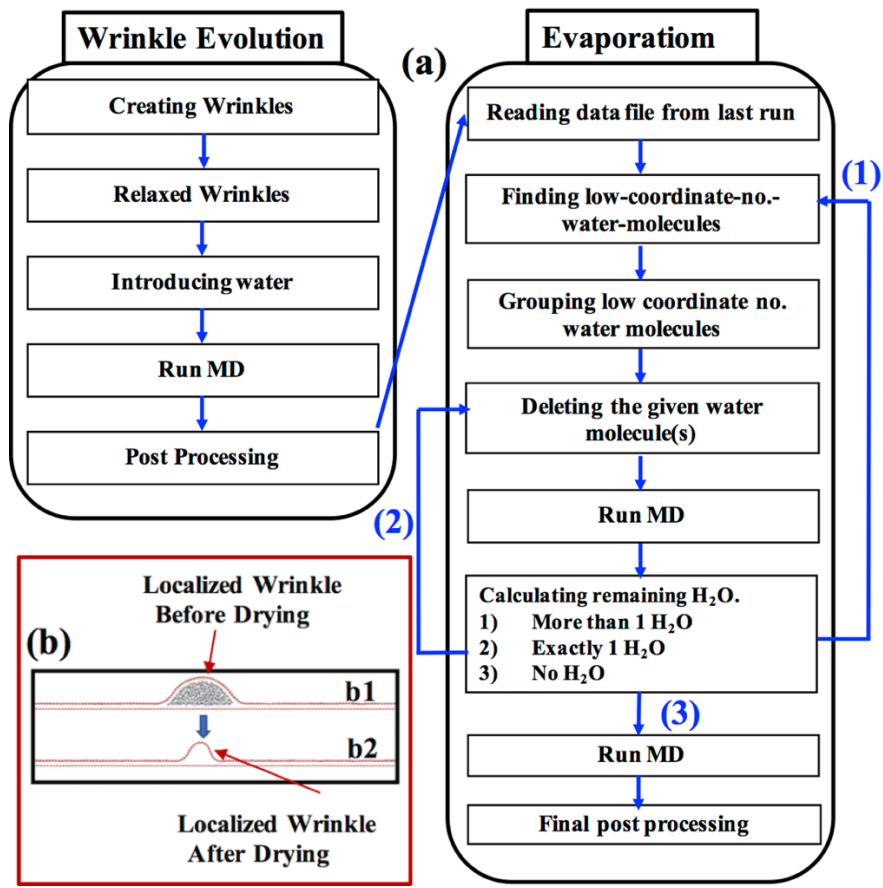

**Figure 3:** (a) Flowchart showing the algorithm used for the Molecular Dynamics (MD) simulation for wrinkle evolution and water evaporation. (b) Snapshot of complete water evaporation.

In *Phase II*, we studied how wrinkle collapses due to the drying of water. The 'Evaporation' part of Figure 3 summarizes the methodology for studying the drying phase. For drying modeling, we started with the last configuration of the system in the first phase, e.g., schematic in Figure 3b1. In our LAMMPS script, the water molecules at the outer periphery were identified. The identification process was accomplished by locating and then grouping those water molecules having the less coordination numbers as compared to their counterparts inside the water-body. The identified surface water molecules were deleted, followed by running the MD simulation for 2000 steps. The same loop was performed again (water deletion and MD simulation) until all water molecules were deleted.

## 2.2 Density Functional Theory (DFT)

In *Phase III,* the DFT based analyses were performed on one frame from each case considered from Phase I, as shown in Figure 11. In this phase, we tried to understand the impact of geometrical parameters of WGNR on their fundamental electronic properties. We implemented first-principles DFT with plane-wave basis sets and pseudopotentials to describe the electron-ion interactions as implemented in Quantum Espresso (QE) [53]. All calculations were done using the projector augmented wave (PAW) pseudopotentials and the Perdew–Burke–Ernzerhof (PBE) exchange-correlation functional [54]. The plane-wave basis sets were used with a plane-wave cutoff energy of 100 eV, and a kinetic energy cutoff for augmentation charges of 400 eV. The convergence threshold of Kohn-Sham equations was set to 1$e$-06. The gamma-centered *k*-point sampling grids, obtained using the Monkhorst–Pack method, were 8x8x8 with a unit offset for the graphene unit cell. The valence electrons contain *s* and *p* orbitals for carbon. Before DFT calculations, all atoms in the cell, as well as the lattice dimensions and angles, were relaxed to the equilibrium configurations by using MD. For consistency, only one case corresponding to two water ($H_2O$) layers was studied from each considered IAW, i.e., 6°, 11°, 16°, 21° (Figure 11). Furthermore, the systems from all four cases were trimmed down to 120 atoms to eliminate the system size dependency, and conform to the available computational resources. For the band structure calculations, the symbols and coordinates of the high-symmetry points in the first Brillouin zone of the crystals were taken from Y Hinuma et al.[55]. MATLAB and VESTA codes were used for the post-processing of the results.

**List of Abbreviation**

FG - Flat Graphene
GDW - Graphene with Distributed Wrinkles
GLW - Graphene with Localized Wrinkle
GNR – Graphene Nano-ribbons
WGNR – Wrinkled Graphene Nanoribbons
IAW - Initial Angle of Wrinkling ($\theta_{IAW}$)
FAW - Final Angle of Wrinkle ($\theta_{FAW}$)
PBC - Periodic Boundary Conditions
CR - Carbon Ratio
AR - Aspect Ratio

# 3. Results and Discussions

We discuss our results for three phases – (I) The formation and evolution of wrinkles for no-water, and water inside GDW and supported FG (without drying), (II) The collapse of the localized wrinkle during water drying, and (III) The electronic structure of free-standing WGNR for four different geometries.

## 3.1 Phase I – The formation and evolution of wrinkles due to diffused water

As discussed in section 1 (introduction), during the synthesis [56] and/or transfer[57], graphene can inherently have distributed wrinkles, and they tend to coalescence to form a localized structure, i.e., from 'Graphene with Distributed Wrinkles (GDW)' to 'Graphene with Localized Wrinkle (GLW)'. For the simplicity of analysis, we considered here GDW only among other available morphologies (Figure 2b), i.e., ripples, crumples, folds, etc. First, we analyzed how GDW transforms into GLW when there is no diffused water (Figure 2b). Figure 4a shows the case of distributed wrinkles for the representative case of $\theta_{IAW} = 21°$ and CR=1.21. As mentioned earlier in section 2, Carbon Ratio (CR) is the ratio of C atoms in the upper sheet with wrinkles and the underlying flat sheet (see Table S1 of SI). The video related to this phenomenon is shown in the SI (Movie_Figure4a.mp4). The distributed wrinkles on same-side amalgamate to form a localized wrinkle, i.e., GDW transforms to GLW. In crystalline solids, same-sign dislocations repel according to Frank's rule [58]. However, in van der Waals (vdW) layered structures, such as graphene, same-sign (same-side) winkles attract each other to form a localized wrinkle. A similar observation was reported for the MoS$_2$ bilayer system [59]. As shown in Figure 4a, for $\theta_{IAW} = 21°$, first, the initially distributed wrinkles transform into intermediate wrinkle localization and finally single 'localized wrinkle'. During the intermediate steps (step = 10000, 25000), initially distributed wrinkles form three followed by two wrinkles and, finally, one 'localized wrinkle'. There is a drop in potential energy during the transformation of three to two and, finally, one wrinkle. In SI, Figures S1, S2, S3 show the evolution of initial wrinkle for three other cases – (1) $\theta_{IAW} = 6°$, CR = 1.017; (2) $\theta_{IAW} = 11°$, CR = 1.053; (3) $\theta_{IAW} = 16°$, CR = 1.11. As shown in Figure S1, when the $\theta_{IAW}$ is less (e.g., $\theta_{IAW} = 6°$), and CR is close to 1, initially distributed wrinkles do not converge to single localized wrinkle, but tries to transform into a highly strained flat sheet.

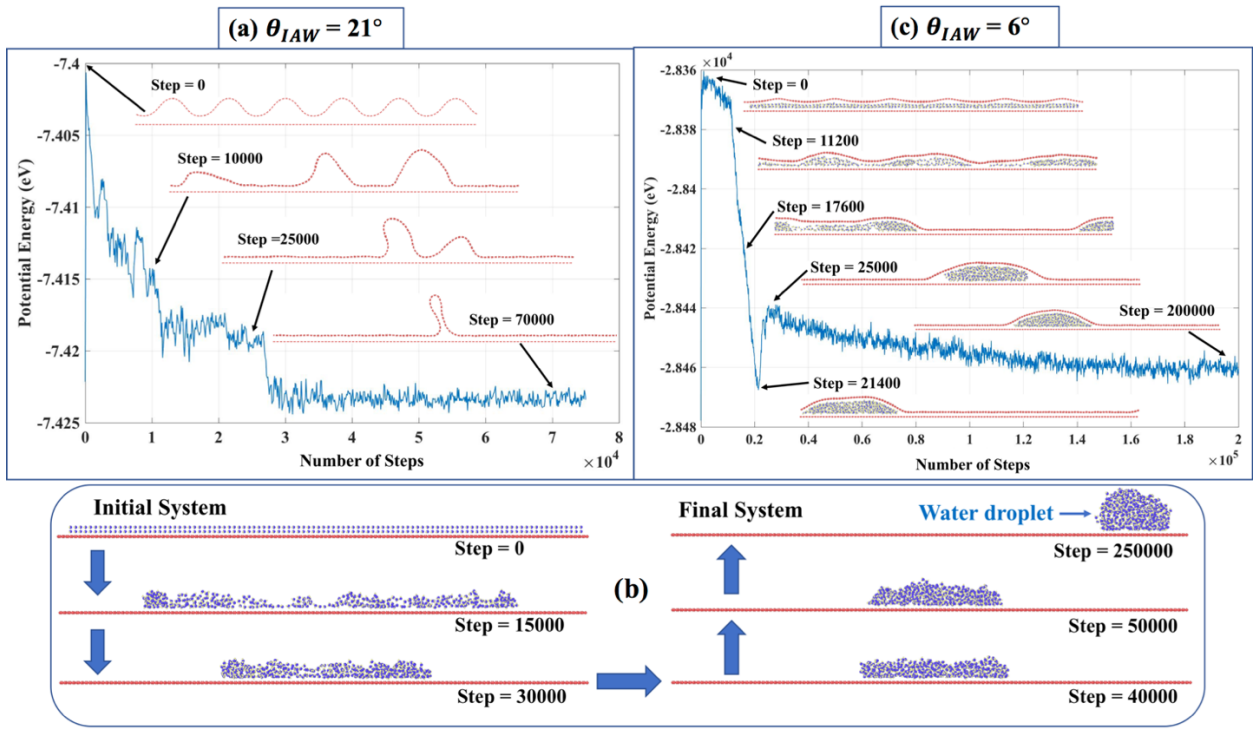

**Figure 4:** (a) Potential energy variation during the wrinkle amalgamation for $\theta_{IAW}$ =21° (no water case), (b) Formation of the water droplet on graphene, (c) Potential energy variation during wrinkle amalgamation ($\theta_{IAW} = 6°$ for two water layer case.)

Figure 4b represents a case-study where water molecules are initially placed on monolayer FG with no graphene sheet on top. The SI contains the related video (Movie_Figure4b.mp4). The MD simulation shows that water molecules coalescence to form a water droplet because of the hydrophobicity of graphene[3] with a contact angle of 125°, which matches with the literature [60]. In Figure 4c, we considered a representative case of $\theta_{IAW} = 6°$ (two-layer water) to discuss how diffused water molecules inside GDW and FG influence the wrinkle evolution. The SI contains the related video (Movie_Figure4c.mp4). The presence of diffused water causes 'competition' between two phenomena – (i) wrinkle amalgamation, and (ii) droplet formation. The final localized wrinkle configuration is largely determined by the dominant phenomenon. Figure S1 shows that for GDW with $\theta_{IAW} = 6°$, initial wrinkles tend to merge together to form a flat sheet. Figure 4b shows the water droplet formation starting from the initial water layer. When these phenomena, i.e., wrinkling formation and droplet formation interact with each other, the droplet formation process dominates. However, the final droplet shape in this case (Figure 4c) is not the same as in Figure 4b. Because in this case, the upper graphene sheet tries to compress the water droplet, as observed by S. McKenzie et al. [61]. However, since the droplet formation phenomenon is dominant here, the final wrinkle configuration is primarily determined by the water droplet.

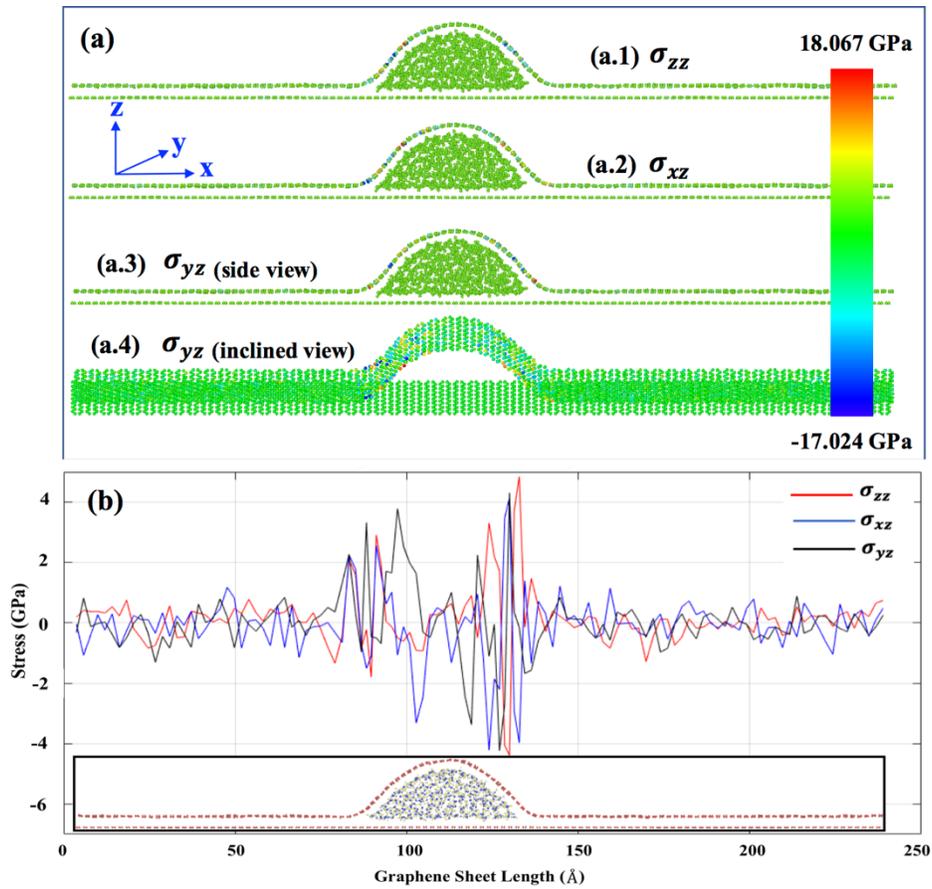

**Figure 5.** (a) Cauchy stress field distribution in the localized wrinkle graphene system for a particular time frame – [a.1] $\sigma_{zz}$, [a.2] $\sigma_{xz}$, [a.3] $\sigma_{yz}$ - side view, [a.4] $\sigma_{yz}$ - inclined view, (b) Stress distribution across the top wrinkled graphene sheet. Stresses are averaged upon the whole line of atoms along the y-axis.

Figures S4, S5, S6, S7 show the final wrinkle configuration of graphene starting from the initial wrinkle angle of $\theta_{IAW} = 6°$, 11°, 16°, and 21° respectively for 0-, 2-, 4-, and 6-layer water cases. The stress generated during the wrinkle formation is shown here. We considered a representative case of $\theta_{IAW} = 11°$

for the 2-layer water case (Figures S5b) for analyzing the stress generation during the wrinkle formation. Figure 5 shows the $z$-component of the Cauchy stress. Considering the stress-field in Figure 5a, we note that atoms in the flat regions and at the top of the wrinkle have no $z$-component stress. Only the sidewalls, i.e., the base of the localized wrinkle, are stressed. Figure 5b shows the distribution of the $z$-component of stress across the GLW (Graphene with Localized Wrinkle). Here, stresses are averaged upon the whole line of atoms along the periodic $y$-axis. We note that for all three stress cases ($\sigma_{zz}, \sigma_{xz}, \sigma_{yz}$), the stress is high in the sidewalls (the base) of the wrinkle as expected [62]. The maximum individual atomic stress, observed in the wrinkle, varies from around -17 to 18 GPa (Figure 5a). When averaged around the whole line of atoms along the periodic $y$-axis, the maximum magnitude of stress is around 4 GPa (Figure 5b). In summary, stress generation in graphene during the wrinkle formation is much lower than the stress that can cause plastic deformation [59,60]. Hence, water diffusion-induced wrinkle formation in graphene does not cause in-plane lattice-strain high enough to break the bonds and cause plasticity.

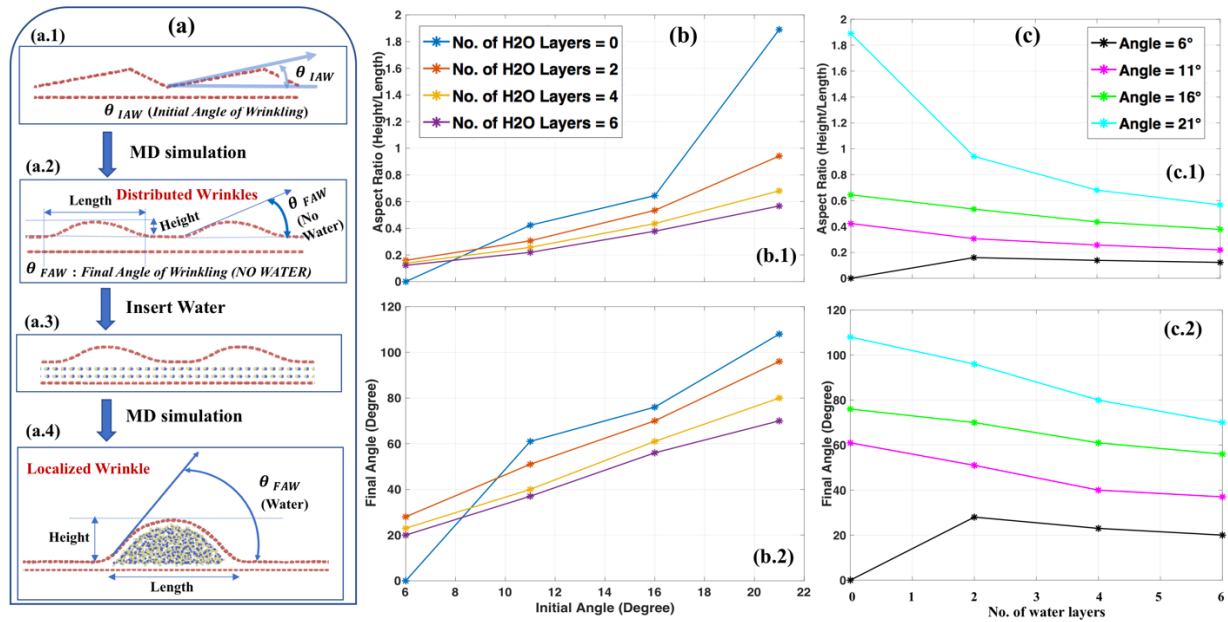

**Figure 6.** (a) Wrinkle configurations at different stages. [a.1] Model considered at the beginning of the simulation showing the 'Initial Angle of Wrinkle ($\theta_{IAW}$)'. [a.2] Wrinkle configuration obtained after MD simulation of Figure a.1 with specific height, length, and Final Angle of Water ($\theta_{FAW}$) of the wrinkled structure (with no water). [a.3] Insertion of water molecules in the minimized wrinkled structure obtained in Figure a.2. [a.4] Final configuration (after MD simulation of Figure a.3) with final height, length, and $\theta_{FAW}$ of the wrinkled structure (with water). (b) Variation of the [b.1] aspect ratio (height/length), and [b.2] final angle ($\theta_{FAW}$) of localized wrinkle obtained from different initial angles ($\theta_{IAW}$). At each $\theta_{IAW}$, we varied the number of water molecules inside the graphene bilayer. [c] The aspect ratio (AR) and final angle variation w.r.t. the no. of water layers.

As shown in Figures S4, S5, S6, S7, irrespective of the number of the diffused water layers (i.e., 2-, 4-, or 6-layer), the pattern of the final configuration is the same for all water layer cases. We define two new parameters, as shown in Figure 6a – (1) *Aspect Ratio (AR)* as the ratio of the height of the wrinkle to its base length, and (2) '*Final Angle of Wrinkle*, $\theta_{FAW}$'. If *AR = 0*, it means the sheet is flat (height = 0). The higher the value of *AR*, the lesser the radius of the curvature of wrinkle-vertex. Moreover, there is a directly proportional relationship between AR and $\theta_{FAW}$. Figure 6b,c show how $AR$ and $\theta_{FAW}$ vary with $\theta_{IAW}$ (*initial angle of water*) for the differing number of the water layers. When there is no diffused water (Figures S4a, S5a, S6a, S7a), as $\theta_{IAW}$ increases from 6° to 21°, the final localized wrinkle becomes sharper. Therefore, the blue curve (*no water case*) in Figure 6b shows that the $AR$ and $\theta_{FAW}$ increase as $\theta_{IAW}$ increases. As mentioned before, the presence of diffused water starts the complex interplay between the wrinkle amalgamation and droplet formation [65]. The final shape is the result of the interaction between two phenomena. For 2-, 4-, or 6-layer diffused water cases (Figures S4b-d, S5b-d, S6b-d, S7b-d), $AR$ and $\theta_{FAW}$ linearly increases as $\theta_{IAW}$ increases from 6° to 21°. As shown in Figure S4a, for $\theta_{IAW} = 6°$ case, the graphene sheet with initial distributed wrinkle flattens ($AR = 0$, $\theta_{FAW} = 0°$ in Figure 6b.1) for no water case. With the insertion of 2-, 4-, or 6-layer diffused water (Figure S4b-d), the droplet and wrinkle size increases. However, the decrease in $AR$ for different water-layer case for $\theta_{IAW} = 6°$ is not significant (*black* line in Figure 6c.1) as the length and height of the wrinkle increase almost proportionately. For the $\theta_{IAW} = 21°$ case (Figure S7), we have the spiker localized wrinkle for no-water case (Figure S7a). The $AR$, in this case, is high (*blue* line in Figure 6b.1), since the base length of wrinkle is very short as compared to the height of the wrinkle. With the increase in diffused water amount, the base length of the wrinkle keeps increasing while the height is almost the same (Figure S7b-d). Thus, $AR$ and $\theta_{FAW}$ values drop significantly (*cyan curve* in Figure 6c).

In Phase I, it is observed that for all cases considered, distributed wrinkles evolve into a localized wrinkle. The way it appears is either the intra-attraction between water molecules to make one droplet, and the vdW attraction between the wrinkle and water molecules that "drag" the graphene along with water droplet [66]. Another possibility is the vdW attraction between the upper graphene sheet and graphene substrate, causing that drag. The probability of the vdW attraction between walls of the same wrinkle, during evolution, are slim to none given the wrinkle wavelength. Furthermore, it may be a combination of all these factors, which may be the function of temporal and spatial variables as well. Though the detailed analysis is beyond the scope of this study, in Phase I, we tried to get an insight into the underlying mechanisms into the wrinkling formation and evolution kinetics.

### 3.2 Phase II: The dyamics of the localized wrinkle due to water evaporation/drying

As shown in Figure 1a, the diffused water evaporates, which changes the structure ($AR$ and $\theta_{FAW}$) of the localized wrinkle. We consider the final localized wrinkle structure for $\theta_{IAW} = 11°$ case with 2-layer water diffusion (Figure S5b) and study the variation in wrinkle configurations upon evaporation. In Figure 7a, we note the sequence of steps of water evaporation and the change in wrinkle configuration. The SI contains the related video (Movie_FigureX7a.mp4). Figure 7b shows the corresponding change in potential energy. We performed the comparative analysis in this plot (as any other potential energy plot in this work), with the least interest in absolute values. When water starts evaporating, the potential energy drops drastically as there is a significant perturbation in the system. Upon drying, wrinkle tends to be spikier, which is analogous to the experimental observation shown in Figure 1. Similarly, we studied the water evaporation and wrinkle formation for other initial angle cases ($\theta_{IAW} = 6°, 16°, 21°$). Figure S8 shows the wrinkle for 6-layer diffused water case for $\theta_{IAW} = 21°$. Upon drying, we notice the spikier wrinkle. The SI contains the related video (Movie_FigureS8.mp4).

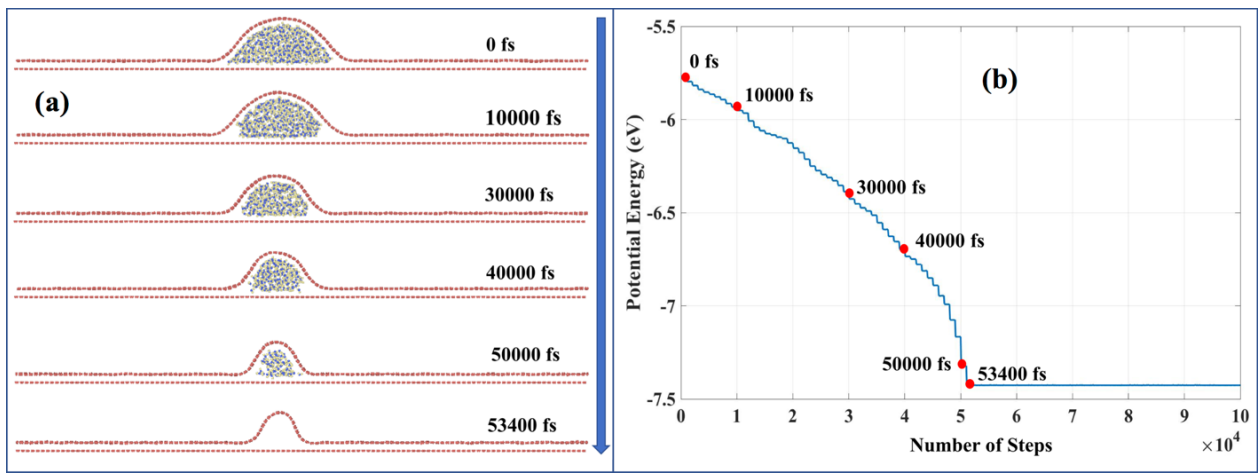

**Figure 7.** (a) The depiction of the water drying process. The system is run by NVE ensemble at 300 K with time step of 1 fs. (b) Potential energy variation during the process.

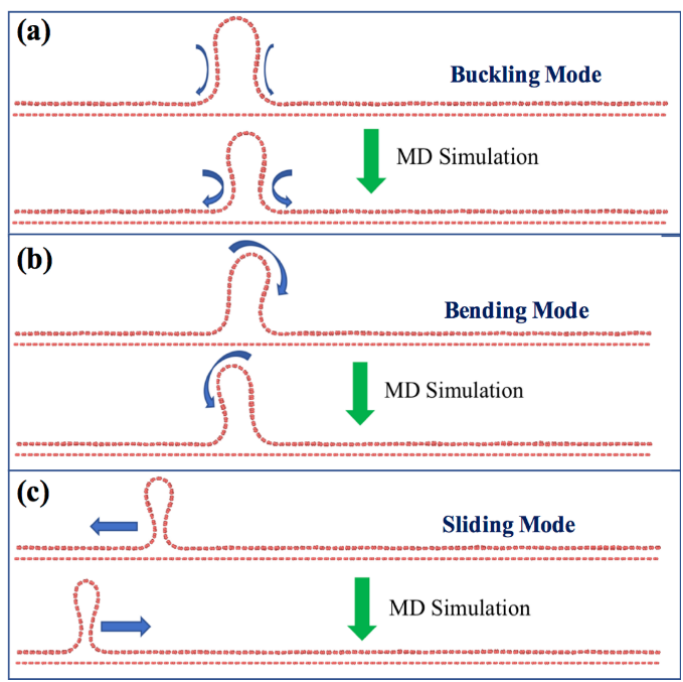

**Figure 8.** After complete drying, the final localized wrinkled structure can undergo three different modes – (a) Buckling, (b) Bending, and (c) Sliding.

After the complete evaporation, the localized wrinkle undergoes structural changes, which results in three kinds of fundamental movements – (1) *Buckling Mode* (Figure 8a): The two sides of the wrinkle buckles and reduces the base-length of the wrinkle. However, the height of the wrinkle remains the same (*AR* increases). (2) *Bending Mode* (Figure 8b): The base length and height of the wrinkle remain the same (*AR = constant*). However, the wrinkle bends sideways. (3) *Sliding Mode* (Figure 8c): In this case, after the complete evaporation, the final localized wrinkle does not remain static at a particular location. Instead, it keeps moving left and/or right. The sliding mode is more or less present in all cases beyond $\theta_{IAW} = 6°$, i.e., after complete evaporation, wrinkle moves sideways. The bending and buckling modes dominate as the initial angle increases. Figure S8 of SI shows the evaporation for a 6-layer water diffusion case. Soon after the complete drying, we notice the buckling at the side of wrinkle. The buckling mode is accompanied by sliding and bending mode (See the related video in SI - Movie_FigureS8.mp4). The bending mode is less dominant for low $\theta_{IAW}$ cases (6°, 11°) when the height of the wrinkle is lesser.

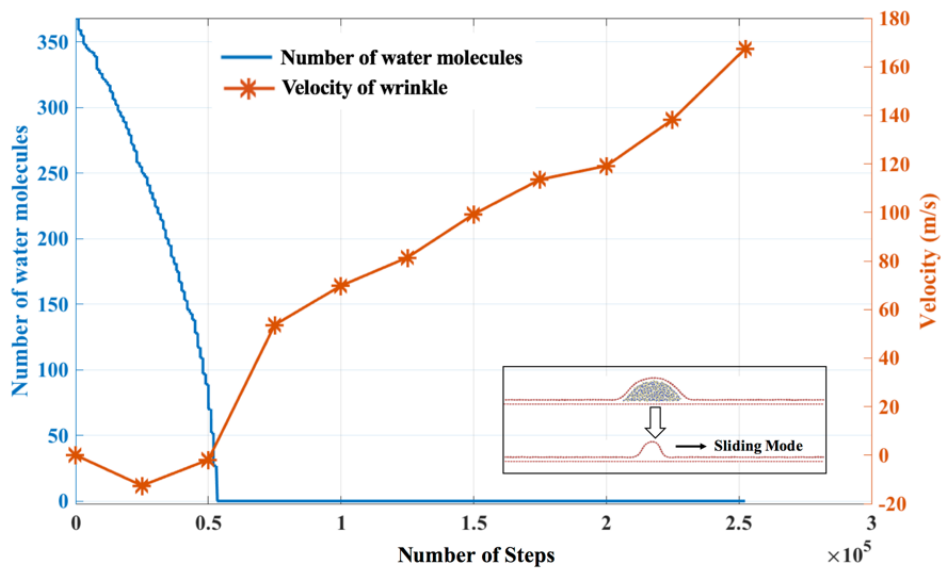

**Figure 9.** Number of water molecules during the drying process and velocity of the wrinkle (averaged over 50,000 steps).

A critical study is needed to analyze which mode (sliding, bending, buckling) is more dominant at different stages of evaporation, given various initial angles, to tune the wrinkle according to the need as targeted by other methods [43]. We attempted to get insights into the domination of sliding mode during drying. As mentioned before, for a lower initial angle (e.g., $\theta_{IAW} = 11°$), the sliding mode is more prominent. Figure 9 shows the velocity of the wrinkle as a function of water content. The 'wrinkle-velocity' is computed with reference to the atom present at the crest of the wrinkle. The identity of the atom located at the crest is changing as wrinkle is moving under the sliding mode. We utilized the 'in-built' python-scripting facility in visualization package OVITO [67] to identify the atomic position at the crest for a different time frame and monitor its movement to compute the velocity. In Figure S9 of the SI, we plotted the number of water molecules present and the corresponding 'wrinkle-velocity'. Because of the atomic scale movement, there is a lot of noise in the velocity data. Here, positive data represents velocity when wrinkle moves towards the positive *x*-axis, while negative represents when wrinkle moves along the negative *x*-axis. In Figure 9, the velocity is averaged over every 50,000 steps. We note that as long as water molecules are diffused inside the wrinkle, the speed ranges around zero (until 0.5x1$e$5 MD steps). However, once the water molecules are removed completely, then not only velocity jumps to a higher value of magnitude, but it also shows a positive gradient with MD time steps, and the wrinkle prefers to move towards the positive *x*-axis. This 'wrinkle-velocity' pattern indicates that the water molecules diffused under the wrinkle behaves as 'wrinkle-anchor' (speed breaker). After the water molecules are completely evaporated, the sideways movement of the wrinkle dominates because the 'wrinkle-anchor' is removed. We hypothesize that the direction of the wrinkle movement after the complete evaporation is decided by the complex interplay of three different modes at the moment when the "last-set" of water-molecules is evaporated. However, further critical analysis is needed to validate this proposition.

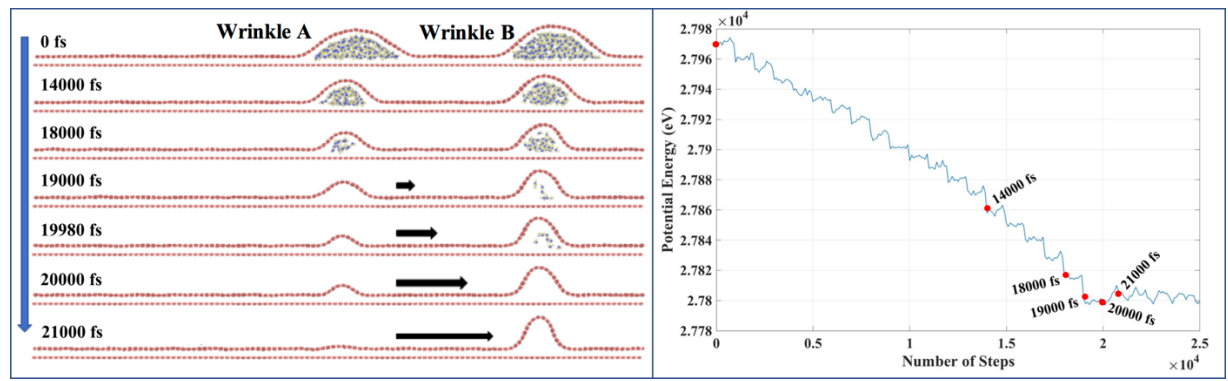

**Figure 10.** (a) Wrinkle amalgamation with water content. The system is run with an NVE ensemble, time step of 1 fs at 300 K. (b) Variation of the potential energy during the wrinkling amalgamation.

To understand further the 'wrinkle-anchor' phenomenon by the diffused water molecules inside the wrinkle, we considered another set-up, as shown in Figure 10. We artificially created a scenario where two wrinkles are placed apart, and the different number of water molecules are diffused inside. Different snap-shots are taken until the system has reached the global minima. SI contains the related video (Movie_FigureX10a.mp4). Wrinkle-A and Wrinkle-B do not amalgamate until one of them lost all the water molecules diffused underneath. As shown in Figure 10, until time = 18,000 fs, the vdW interaction between the diffused water and the substrate acts as an 'anchor support' to provide the spatial rigidity to the wrinkle. At time = 19,000 fs, Wrinkle-A has the complete evaporation, while for Wrinkle-B, some diffused water molecules are still present. The diffused water molecules act as 'wrinkle-anchor' (speed breaker) for Wrinkle-B. Moreover, the presence of water molecules inside Wrinkle-B also prevents the sliding mode of Wrinkle-A. In fact, instead of sliding mode, here Wrinkle-A starts collapsing after complete evaporation since Wrinkle-B prevents its sideway motion. After complete evaporation for Wrinkle-B (time = 20,000 fs), Wrinkle-A collapses completely and merges with Wrinkle-B. Therefore, unlike single wrinkle cases (Figure 7), when two wrinkles are present, the motion of one wrinkle not only depends on the amount of water diffused underneath it but also on the water underneath the adjacent wrinkle. We hypothesize that in a network of wrinkles in a 2D material, we can control the evolution of the wrinkle-network by controlling the amount of water content trapped under fewer wrinkles than the total number of wrinkles participating in the wrinkle-network. It will be valuable observation [68] if we can control the evolution of the whole wrinkle-network by controlling the water content trapped under just one wrinkle, which may be or may not be the member of the original wrinkle-network.

### 3.3 Phase III – DFT study of the electronic properties of winkled graphene

#### 3.3.1 Crystal orbitals (HOMO and LUMO):

We performed the Molecular Orbital (MO) analysis of WGNR for all four angles ($6°, 11°, 16°, 21°$). We considered the localized wrinkle structure for four-layer of water cases shown in Figure S4c, S5c, S6c, and S7c. Then we performed the complete evaporation. The resulting localized wrinkle after complete evaporation is considered here, excluding the underlying substrate, i.e., free-standing WGNR (Figure 11). We considered the same number of atoms in all the cases, i.e., 120 atoms. The system is optimized with the MD code before performing electronic calculations with DFT. Figure 11 shows the isometric view and Fig. S11 of SI shows the front view. Crystal orbitals are shown by the iso-surfaces with yellow color. A carbon atom has six electrons, out of which two lowest electrons are considered as core electrons, and the next four electrons are considered as valence electrons. These valence electrons participate in the interaction with other atoms in the crystal. Consequently, we are considering these four electrons in the outmost shell in the DFT calculations.

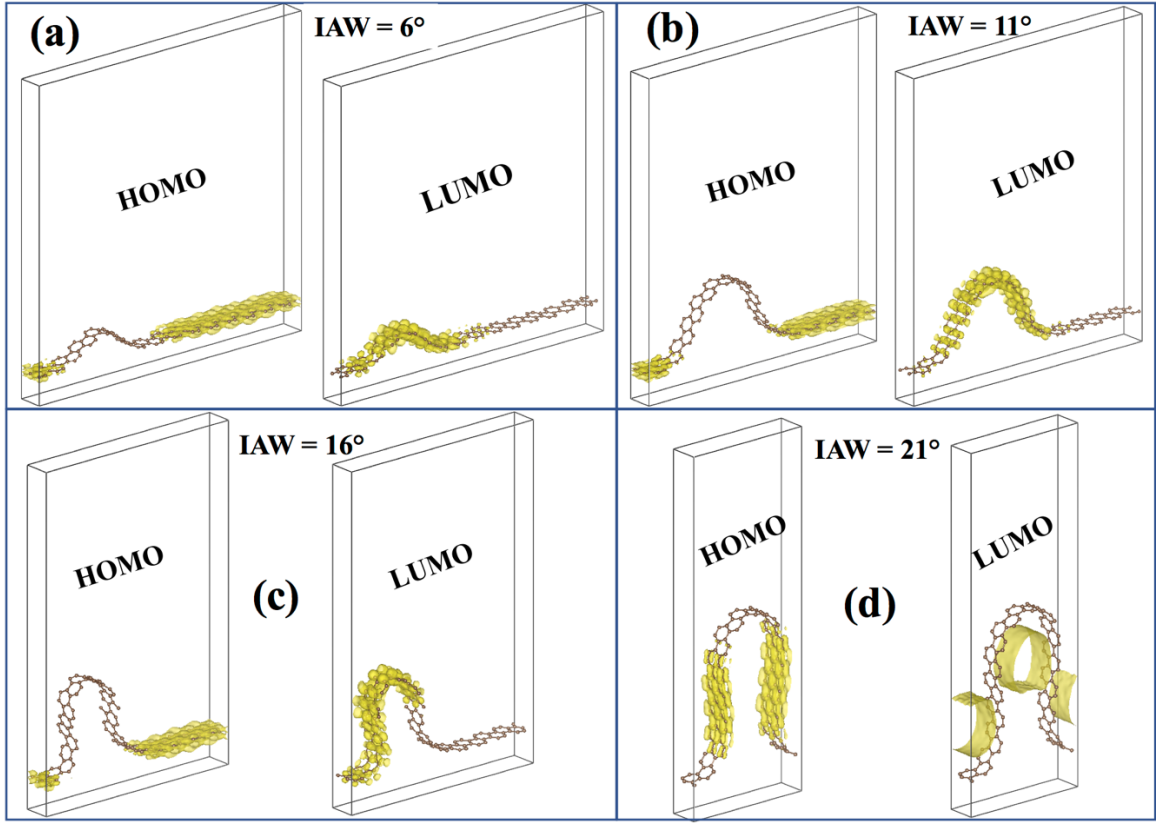

**Figure 11**: Isometric view of HOMO and LUMO for four localized wrinkle obtained after complete evaporation of wrinkled graphene (four water-layer case) with Initial Angle of Wrinkle (IAW) of (a) 6°, (b) 11°, (c) 16°, and (d) 21°.

We observed the Highest Occupied Molecular Orbital (HOMO) at the portion of the WGNR, where the radius of curvature is high, i.e., the flat portion (Figure 11). For IAW of 6°, 11°, 16°, HOMO is observed to be only on the flat part of the GNR. This indicates that lattice distortion, present due to the wrinkle, is increasing the energy level of the molecular orbitals forming up the wrinkled portion of GNR. Hence, the electrons prefer to fill up the molecular orbitals formed by the flat portion of the GNR. The same phenomenon is observed in the sidewalls of the wrinkle for the highest angle considered (21°). In this case, the wrinkle is slender enough so that its sidewalls act like flat graphene and form the molecular orbitals having less energy. Consequently, electrons prefer to stay in the molecular orbitals formed by the sidewalls of the wrinkle instead of the vertex.

The Lowest Unoccupied Molecular Orbital (LUMO) is mostly observed in the portion of the GNR having high curvature (Figure 11). From IAW of 6° to 16°, LUMO can be found in the wrinkled section of the GNR but definitely not in the flat portion. It indicates that if the GNR gains an electron, then it happens by the atoms forming the wrinkle. The effect of curvature plays a dominant role when we increase the IAW beyond 16°. In the case of 21°, we observe LUMO in the form of tunnel shape [69]. The wrinkle is slender enough to confine the lowest unoccupied molecular orbital parallel to its direction. Further investigation is necessary to understand whether it's just the effect of curvature or the presence of HOMO along the sidewalls of the wrinkle or both of these factors contributing to some extent to the formation of "electron-tunnel"-LUMO. This observation proves that not only the presence of wrinkles but its slenderness also plays a crucial role in the generation of "tunnel" LUMO orbital. It can be utilized to control the direction of electron flow in a (semi)conducting system, as observed by J.A Sulpizio et al. in the form of electron flow following a macro hydrodynamic flow pattern, i.e., Poiseuille flow [70].

Similar to these 'electron-tunnels', 'electron-channels' are obtained in the IAW values of 6°, 11°, 16°. The 'electron-channels' is the overlapping of the atomic orbitals of the same energy level forming an "electron-flow-channel" along the direction of wrinkle, which can act as a tunnel to the flow of every extra electron added to the system. Consequently, by controlling the course of these channels, the direction of electron-flow can also be controlled, which has multiple applications in the semiconductor industries. The studies aimed for similar outcomes are available in the literature [71,72]. For IAW of 16°, the asymmetricity is observed for LUMO. As shown in Figure 11c (SI Figure S11c), the wrinkle is bent on one side (left). That causes the LUMO to only form on the curved sidewall completely and partially on the right sidewall. There is a portion of the left sidewall of the wrinkle, where no LUMO or HOMO is observed (SI Figure S11c). Further investigation is necessary to determine the energy level of the orbitals, possessed by the corresponding set of atoms forming the portion of the left sidewall of the wrinkle, where neither LUMO nor HOMO is obtained.

### 3.3.2 Band structure

In Figure 12, band structures are plotted for localized wrinkled graphene structures shown in Figure 11. Each band structure is based upon the system consisting of 240 bands corresponding to 480 electrons constituting the crystal. The direct bandgap is observed for all of the cases. The observed bandgap values, based upon the difference between HOMO and LUMO, are 0.5492, 0.5453, 0.4801, and 0.5266 eV for the localized wrinkled structure corresponding to the IAW cases of 6°, 11°, 16°, 21°, respectively. Figure S10 in SI shows the band structure of the flat pristine graphene. As expected, no bandgap is observed for this case.

The bandgap is observed to be decreasing linearly with IAW (6°, 11°, 16°), but then it increases suddenly for 21°. This can be attributed to the observation made in Figure 11d for the 21° case, i.e., distortion of the HOMO-LUMO. As explained before, because of slenderness, the HOMO is observed on the sidewalls of the wrinkle. Moreover, because of high curvature, LUMO shifted from the tip to under the wrinkle to form an "electron tunnel". The delocalization of the HOMO and LUMO orbital for the highest angle case (21°) may be the reason behind non-linearity in the proportional relationship of IAW and bandgap.

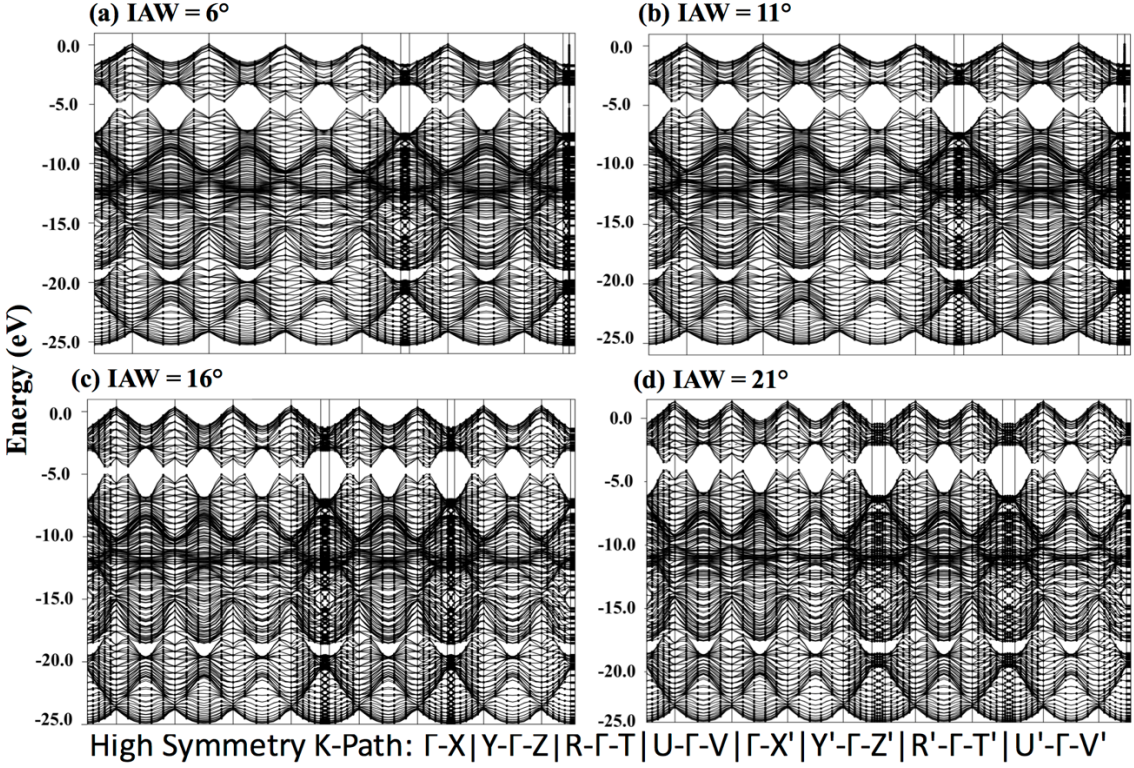

**Figure 12**: Band structure along high symmetry k-points for four localized wrinkle obtained after complete evaporation of wrinkled graphene (four water-layer case) with Initial Angle of Wrinkle (IAW) of (a) 6°, (b) 11°, (c) 16°, and (d) 21°.

### 3.3.3. Density of States (DOS)

In Figure 13, the density of states (DOS) is plotted for four wrinkled graphene structures mentioned in Figure 11. A proportional relationship is observed between Fermi energy and the IAW. As IAW increases (6°, 11°, 16°, 21°), the magnitude of the Fermi energy decreases as -3.3651 eV, -3.2326 eV, -2.9825 eV, -2.0637 eV, respectively. The Fermi energy obtained for flat graphene is -2.3462 eV, which agrees with the values reported in the literature [73]. With the increase in IAW, the final localized wrinkle is spikier. It means that the 'wrinkled zone' is composed of more number of atoms and hence more stretched bonds.

Consequently, the strained lattice increases the energy levels of the localized crystal orbitals, which results in changing the Fermi energy as a function of IAW. In Figure 13, we notice two prominent peaks in DOS plots for all cases (IAW = 11°, 16°, 21°) except for the least IAW case (6°). We have the second peak in all the cases at around -2 eV to 0 eV, and the first peak at around -5 eV except for the least IAW case (6°). Further work can be conducted to find out the first value of IAW, which will generate the first peak in between IAW of 6° to 11° and results in two peaks.

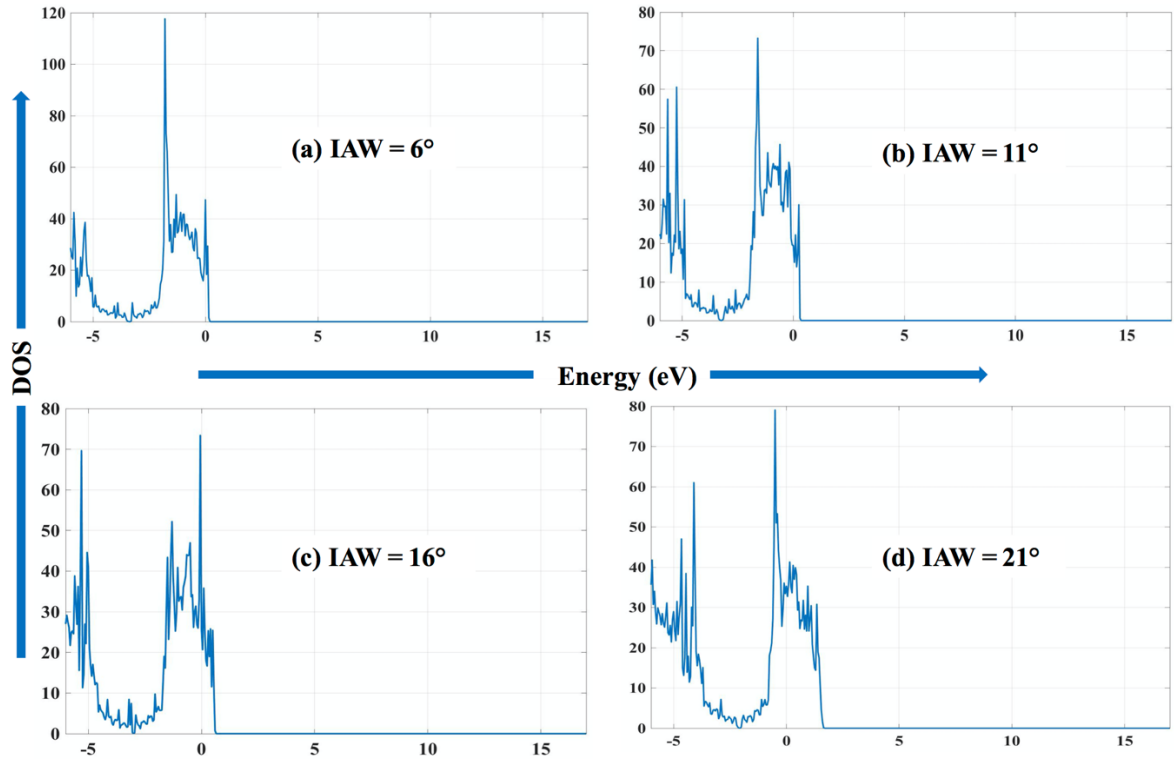

**Figure 13**: Density of states for four localized wrinkle obtained after complete evaporation of wrinkled graphene (four water-layer case) with Initial Angle of Wrinkle (IAW) of (a) 6°, (b) 11°, (c) 16°, and (d) 21°.

## 4. Conclusions

In conclusion, we have performed MD and DFT simulations to geometrically and electronically characterize the wrinkle formation, evolution, and collapse in graphene when water is diffused in between graphene and substrate, and while it is evaporating. Our key findings are summarized below:

- The distributed wrinkles in graphene coalesce to form a 'localized wrinkle', whose configuration significantly depends on the initial pattern of distributed wrinkles and the amount of diffused water present in between the graphene and substrate. In the absence of diffused water, the distributed wrinkles only with the higher initial angle, transform into a spikier localized wrinkle.

- When water is distributed on a graphene sheet, due to hydrophobicity, distributed water molecules condense to form a water droplet. In the presence of diffused water, the final wrinkle configuration is the result of the competition between the water droplet formation process and the localized wrinkle formation process. For the lower initial angle of wrinkle, droplet formation is dominant, resulting in wider (less-spikier) wrinkle. For higher initial wrinkle angle, wrinkle formation dominates, and the final wrinkle is comparatively spiker.

- The drying of water, diffused under the localized wrinkle, changes its configuration, e.g., it becomes spikier. The wrinkle is static until the complete evaporation, and after that, it acquires a motion. In case of diffused water under two localized wrinkles placed side-by-side, even after the complete evaporation under one wrinkle, it remains static until the water diffused under the neighboring wrinkle is completely dried.

- The wrinkle movement is the combination of three fundamental modes – bending, buckling, and sliding. The dominant mode depends on the configuration of the wrinkle., e.g., spikier wrinkle tends to bend. Further investigation is necessary to understand how three different modes influence wrinkle dynamics.

- The stress analysis reveals that the maximum stress is at the base of the wrinkle, and always below its plasticity limit. Therefore, wrinkle formation due to water diffusion does not cause any significant bond-breaking.

- "Electron-tunnel/channel" is observed along the vertex of the WGNR. By controlling the direction of WGNR, we can control the direction of the electron flow. This concept can be used in designing flexible electronics and optoelectronics.

- The direct bandgap is observed for all the wrinkle cases considered. The change in wrinkle geometry also generates lattice strains, which further changes the energy level of the crystal orbitals. The strain in lattice changes the Fermi energies as well as results in Fermi level modulations.

Our study addresses the fundamental questions regarding the wrinkle formation, evolution, and collapse due to water diffusion/evaporation and its importance in flexible electronics.

## Acknowledgement

DD acknowledges the support of the NJIT faculty start-up grant. EH is supported in part by the National Science Foundation award (ECCS-1104870), the Defense University Research Instrumentation Program (FA9550-11-1-0272). We are grateful to the High- Performance Computing (HPC) facilities managed by Academic and Research Computing Systems (ARCS) in the Department of Information Services and Technology (IST) of the New Jersey Institute of Technology (NJIT). Some computations were performed on Kong HPC cluster, managed by ARCS. We acknowledge the support of the Extreme Science and Engineering Discovery Environment (XSEDE) for providing us their computational facilities (Start Up Allocation – DMR170065 & Research Allocation – DMR180013).

## Author Contributions

All authors conceived the project. JK wrote all the codes and performed simulations. All authors discussed the results. JK wrote the manuscript.

## Competing Interests

The authors declare that there are no competing interests.

**Data Availability**

The data reported in this paper is available from the corresponding author upon reasonable request.

**Code Availability**

The pre- and post-processing codes used in this paper are available from the corresponding author upon reasonable request. Restrictions apply to the availability of the simulation codes, which were used under license for this study.

*Nano Lett.*, vol. 15, no. 2, pp. 1302–1308, Feb. 2015.

# SUPPLEMENTARY INFORMATION

## Comprehensive understanding of water-driven graphene wrinkle life-cycle towards applications in flexible electronics: A computational study

Jatin Kashyap[1], Eui-Hyeok Yang[2], Dibakar Datta[*,1]

[1] Department of Mechanical and Industrial Engineering, New Jersey Institute of Technology, Newark, NJ 07103, USA
[2] Department of Mechanical Engineering, Stevens Institute of Technology, Hoboken, NJ 07030, USA

*Corresponding Author (Email: dibakar.datta@njit.edu)

### A. Tables

**Table S1:** Number of carbon atoms in the upper and lower graphene for different $\theta_{IAW}$

|  | $\theta_{IAW} = 6°$ | $\theta_{IAW} = 11°$ | $\theta_{IAW} = 16°$ | $\theta_{IAW} = 21°$ |
|---|---|---|---|---|
| Number of C in the **Upper** 'Graphene with Wrinkles' | 1920 | 1920 | 1920 | 1920 |
| Number of C in **Lower** Flat Graphene | 1888 | 1824 | 1728 | 1600 |
| ***Carbon Atom Ratio (CAR)*** between the Upper Graphene & Lower Graphene | 1.0169 | 1.0526 | 1.1111 | 1.2000 |

**Table S2:** Number of water molecules for different cases

| Water Layer | $\theta_{IAW} = 6°$ | $\theta_{IAW} = 11°$ | $\theta_{IAW} = 16°$ | $\theta_{IAW} = 21°$ |
|---|---|---|---|---|
| 0 | 0 | 0 | 0 | 0 |
| 2 | 384 | 368 | 352 | 320 |
| 4 | 768 | 736 | 704 | 640 |
| 6 | 1152 | 1104 | 1056 | 960 |

## B. Figures

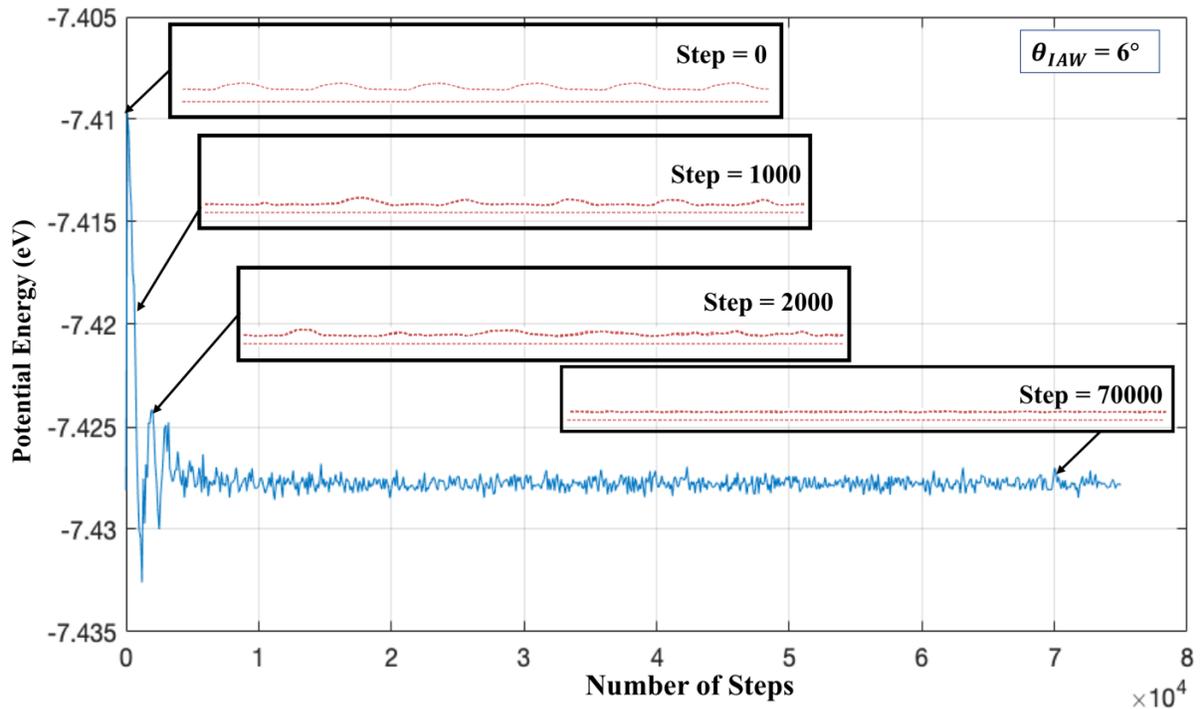

**Figure S1**: Potential Energy variation for initial wrinkle angle of 6° (without water case).

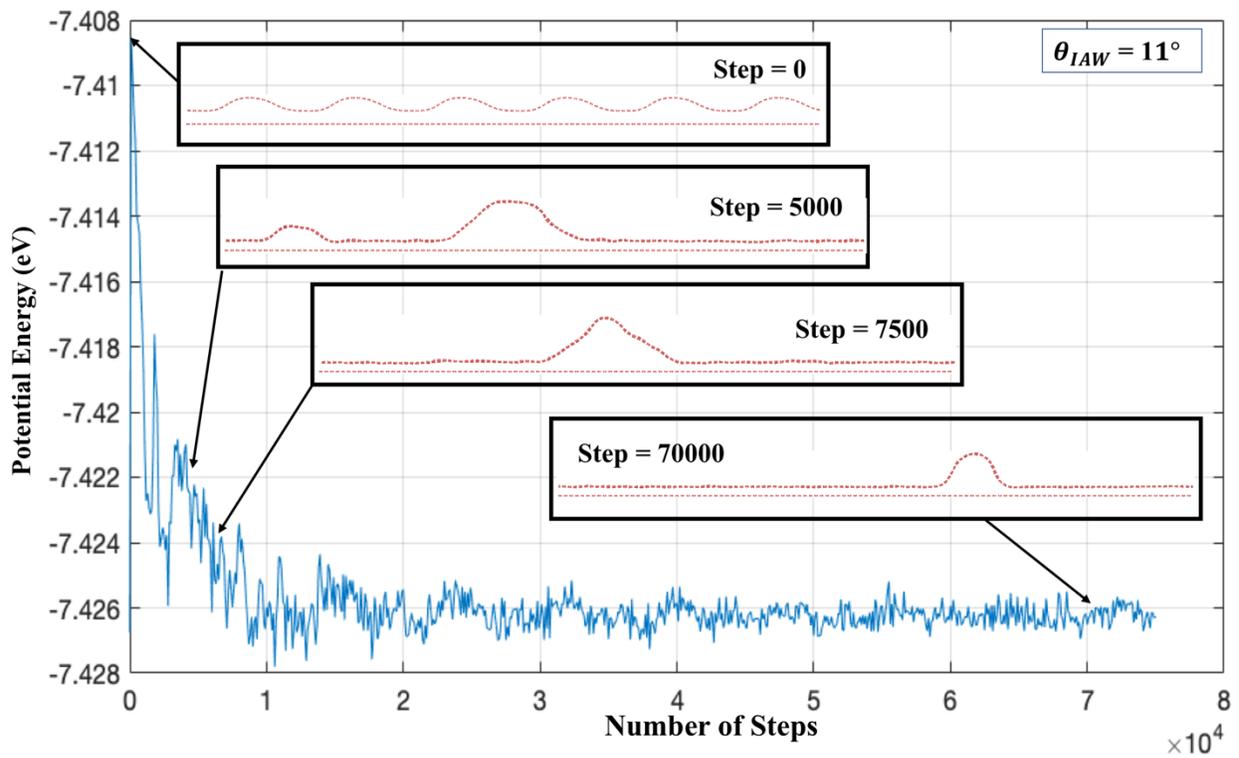

**Figure S2:** Potential Energy variation for initial wrinkle angle of 11° (without water case).

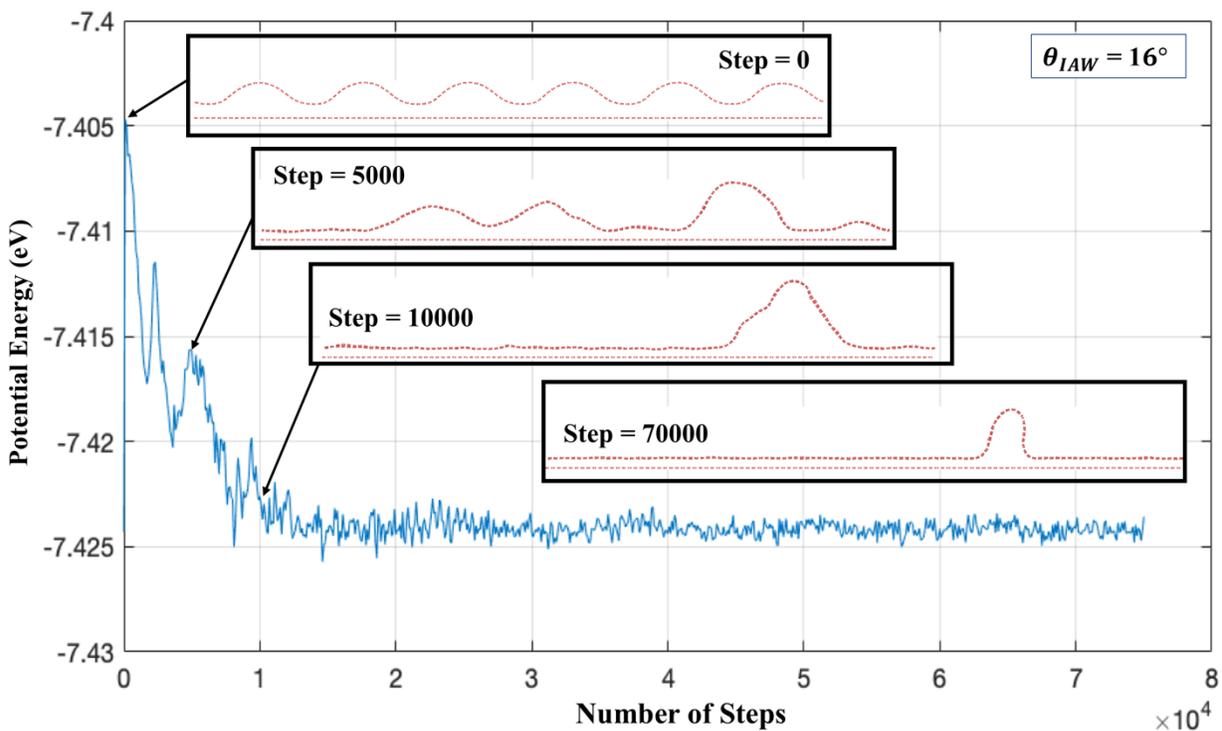

**Figure S3:** Potential Energy variation for initial wrinkle angle of 16° (without water case).

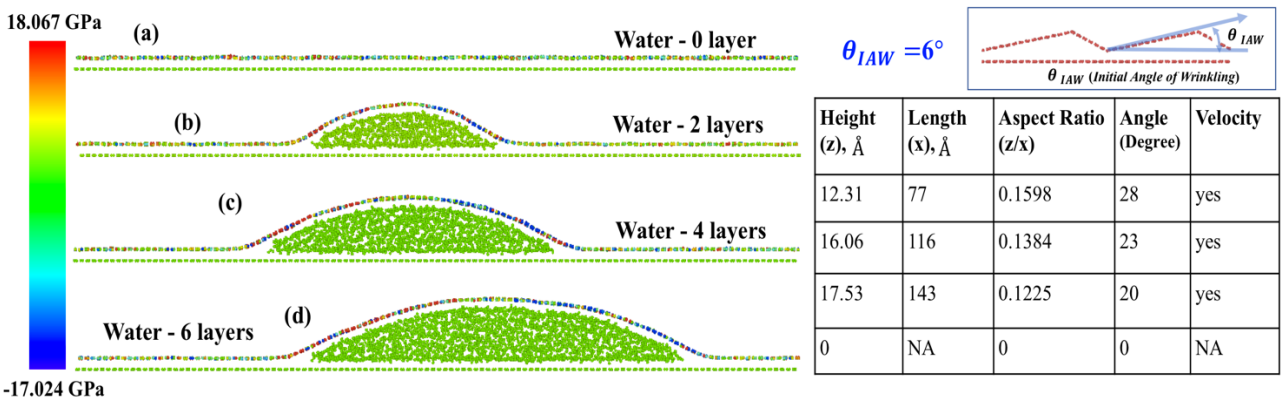

**Figure S4:** Final equilibrium structure with stress profile starting from initial wrinkle angle of 6°.

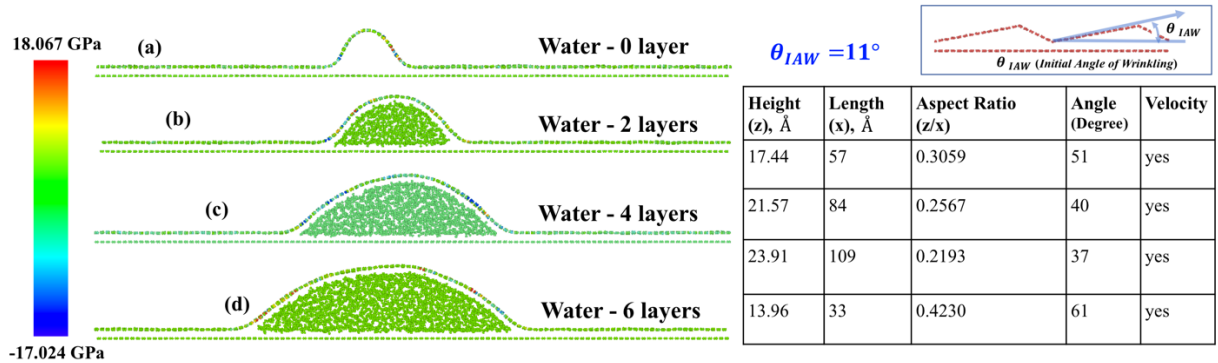

**Figure S5:** Final equilibrium structure with stress profile starting from initial wrinkle angle of 11°.

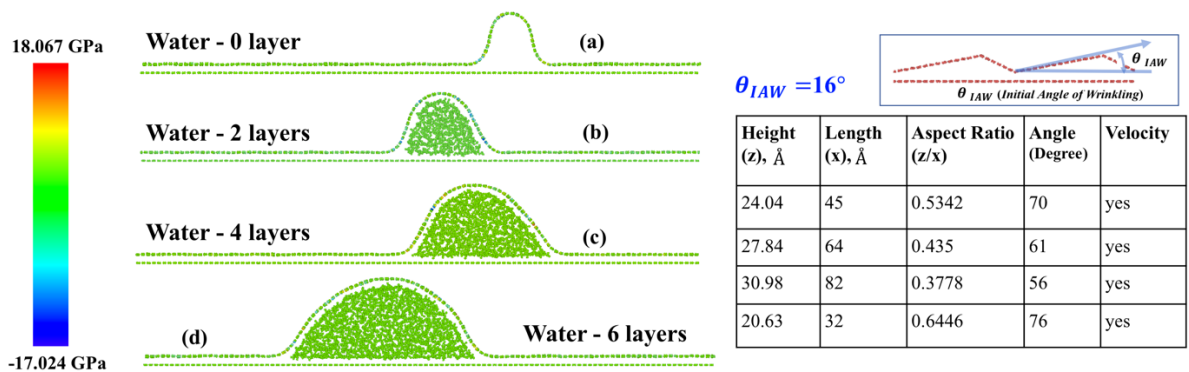

**Figure S6:** Final equilibrium structure with stress profile starting from initial wrinkle angle of 16°.

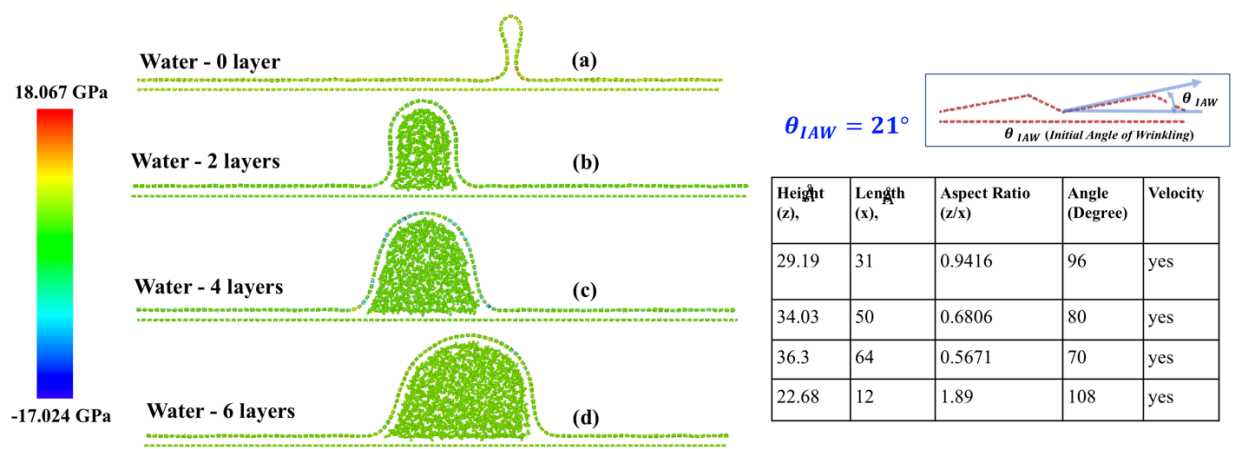

**Figure S7:** Final equilibrium structure with stress profile starting from initial wrinkle angle of 21°.

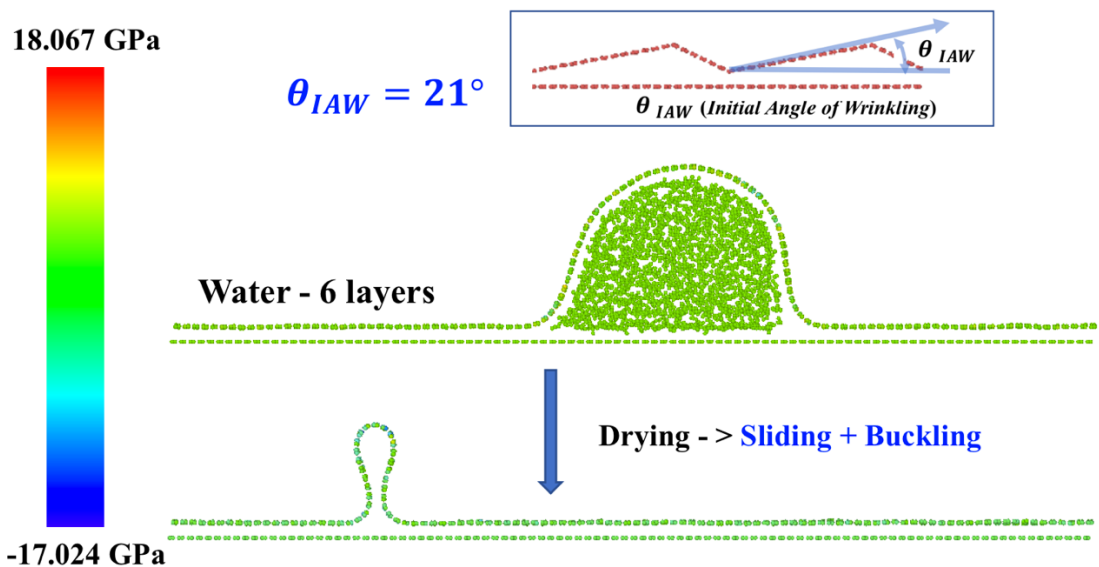

**Figure S8:** Final equilibrium structure starting from initial wrinkle angle of 21°.

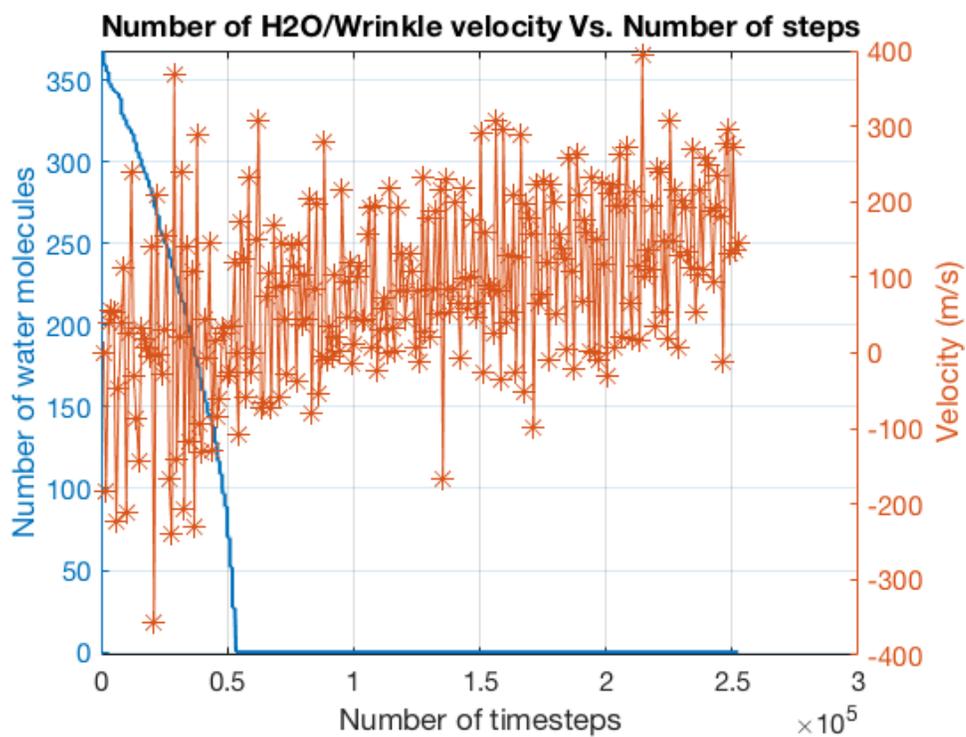

**Figure S9:** Velocity of the wrinkle and the number of water molecules due to evaporation. Here, instantaneous velocity is considered.

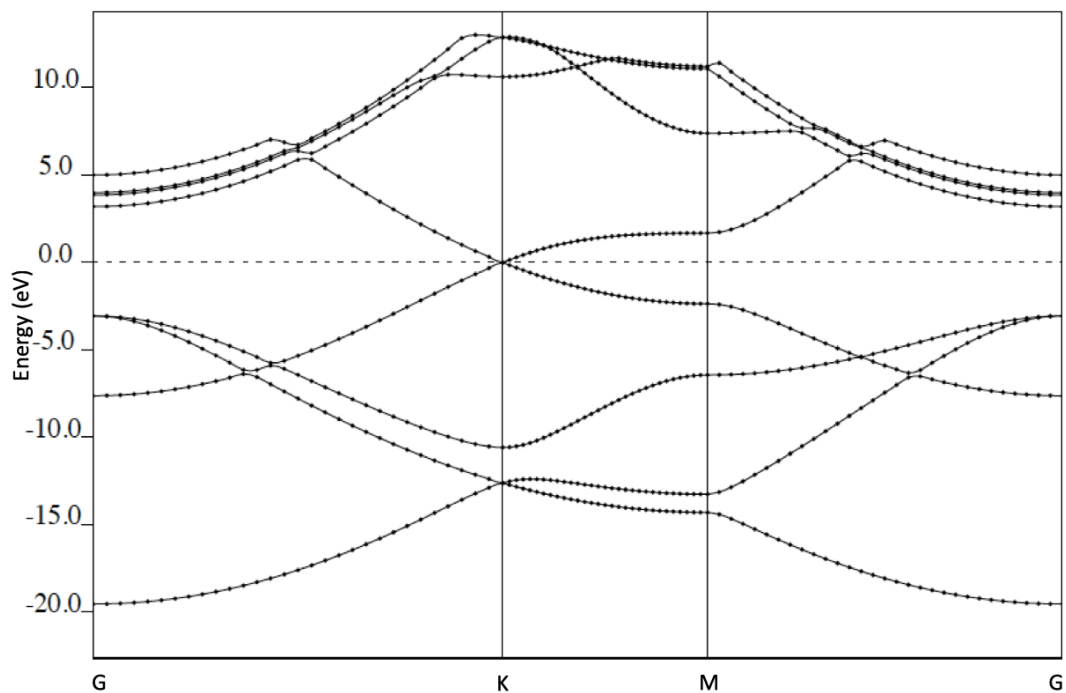

**Figure S10:** Band structure of pristine flat graphene.

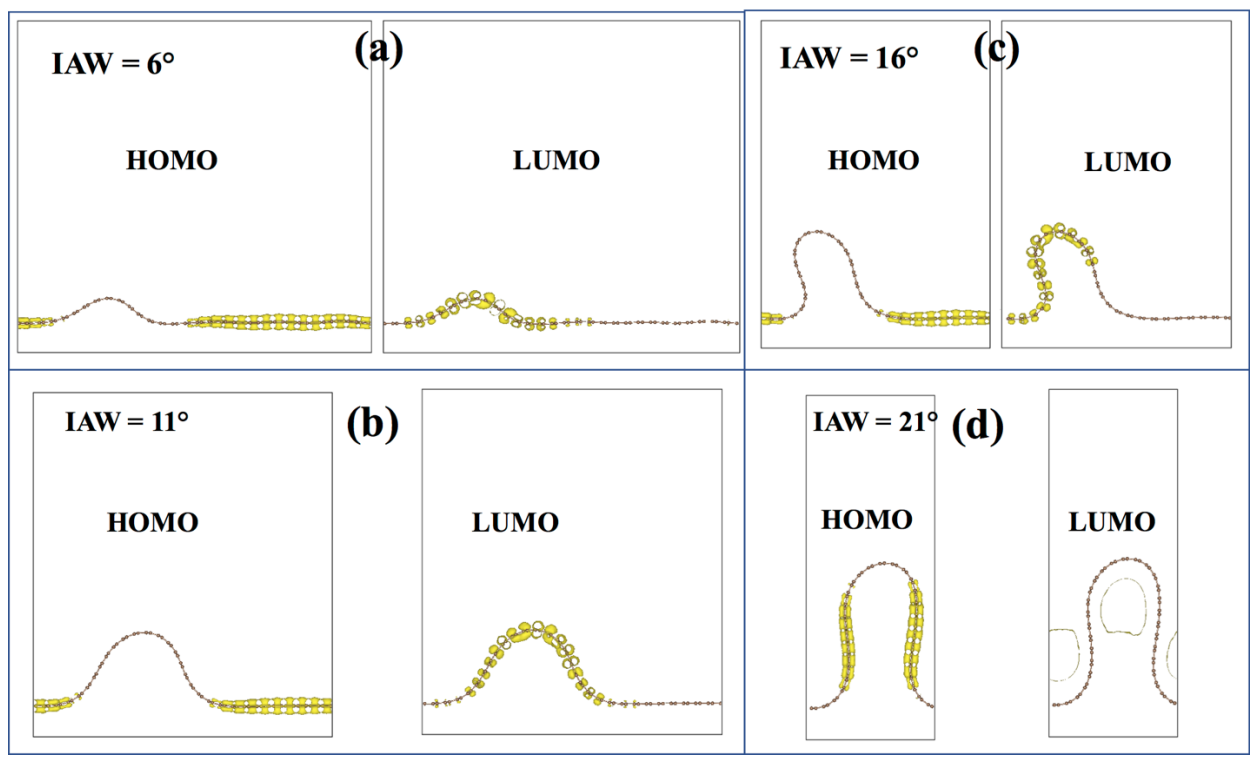

**Figure S11**: Front view of HOMO and LUMO for four localized wrinkle obtained after complete evaporation of wrinkled graphene (four water-layer case) with Initial Angle of Wrinkle (IAW) of (a) 6°, (b) 11°, (c) 16°, and (d) 21°.

### C. Movie Files

| Sl. No. | Movie File Name | Corresponding Figure File Name |
|---|---|---|
| 1 | Movie_Figure4a.mp4 | Figure4a |
| 2 | Movie_Figure4b.mp4 | Figure4b |
| 3 | Movie_Figure4c.mp4 | Figure4c |
| 4 | Movie_Figure7a.mp4 | Figure7a |
| 5 | Movie_Figure10a.mp4 | Figure10a |
| 6 | Movie_FigureS8.mp4 | FigureS8 |

### D. Detail of the stress calculation.

The stress tensor for atom I is given by Equation 1 [1]. Here, subscripts *a* and *b* are substituted with the values of spatial variables associated with given atom I to produce different values of the stress tensor. The first term is a contribution from kinetic energy of atom I. The second term gives the virial contribution associated with intramolecular and intermolecular interactions as given by Equation 2.

$$S_{ab} = -m v_a v_b - W_{ab} \tag{1}$$

Here, m and v are mass and velocity of the atom I.

In Equation 2, the first term is the pairwise energy contribution of atom I, where *n* sums over $N_p$ neighbor atoms. $F_1$, $F_2$ and $r_1$, $r_2$ are forces and positions associated with atoms considered in the pairwise interaction, respectively. The second term contributes to the bonds formed with atom I. Here, the summation is performed over atoms bonded with atom I according to the given potential field. The third, fourth and fifth terms are contributing in the same way as the second term, but the only difference is these terms are considering the angle, dihedral, and improper interactions of the atom I, respectively.

$$W_{ab} = \frac{1}{2}\sum_{n=1}^{N_p}(r_{1_a}F_{1_b} + r_{2_a}F_{2_b}) + \frac{1}{2}\sum_{n=1}^{N_b}(r_{1_a}F_{1_b} + r_{2_a}F_{2_b}) + \frac{1}{3}\sum_{n=1}^{N_a}(r_{1_a}F_{1_b} + r_{2_a}F_{2_b} + r_{3_a}F_{3_b}) +$$

$$+ \frac{1}{4}\sum_{n=1}^{N_d}(r_{1_a}F_{1_b} + r_{2_a}F_{2_b} + r_{3_a}F_{3_b} + r_{4_a}F_{4_b}) +$$

$$+ \frac{1}{4}\sum_{n=1}^{N_i}(r_{1_a}F_{1_b} + r_{2_a}F_{2_b} + r_{3_a}F_{3_b} + r_{4_a}F_{4_b}) + Kspace\,(r_{i_a}, F_{i_b}) + \sum_{n=1}^{N_f} r_{i_a}F_{i_b}$$

$$\tag{2}$$

The sixth term defines the contribution by long range columbic interactions, which are defined in the phase -II of the study due to absence of charged particles, i.e., water molecules. The last term is giving the contribution by constraints used to restrict the movement of carbon atoms of the substrate, i.e, flat graphene (FG).